**Classification:** Physical sciences, Biophysics

# A New Look at Dynamic Force Spectroscopy of Adhesion Bonds.

O.K. Dudko[a], A.E. Filippov[b], J. Klafter[a] and M. Urbakh[a]

*[a]School of Chemistry, Tel Aviv University, 69978 Tel Aviv, Israel.*
*[b]Donetsk Institute for Physics and Engineering of NASU, 83144, Donetsk, Ukraine.*

**Corresponding author:** Prof. Michael Urbakh, Address: School of Chemistry, Tel Aviv University, 69978,Tel Aviv, Israel, E-Mail: urbakh@post.tau.ac.il

Phone: +972-3-6408324 ; Fax: +972-3-6409293

**Manuscript information:** 11 text pages, 3 figures.

**Word and character counts:** 107 words in the abstract, 13,728 characters in the paper.

**Abstract.**

Dynamic force spectroscopy of single molecules is described by a model which predicts a distribution of rupture forces, the corresponding mean rupture force and variance, all amenable to experimental tests. The distribution has a pronounced asymmetry which has recently been observed experimentally. The mean rupture force follows a $(\ln V)^{2/3}$ dependence on the pulling velocity $V$ and differs from earlier predictions. Interestingly, at low pulling velocities a rebinding process is obtained whose signature is an intermittent behavior of the spring force which delays the rupture. An extension to include conformational changes of the adhesion complex is proposed which leads to the possibility of bimodal distributions of rupture forces.

Single molecule spectroscopy is by now an established approach which can report on distributions of molecular properties and can provide kinetic information on conformational changes such as folding and unfolding of molecules without the "scrambling" that occurs due to ensemble averaging [1]. Such information could be valuable in particular for biomolecules, where rare events might have functional significance, but which can be masked in an ensemble approach. Dynamic force spectroscopy (DFS) has been introduced as a spectroscopic tool to probe the complex relationships between "force-lifetime-and chemistry" in single molecules bound in an adhesion complex [2,3] and to reveal details of molecular scale energy landscapes and adhesion strength.

The rupture force in DFS is quantified by the maximum extension of a spring, the linker, which is followed by a rapid recoil of the spring to its rest position. This resembles the stick-to-slip transition in studies on friction. The unbinding process of a single molecule is studied one molecule at a time, which means that one measures a collection of independent random rupture events. This type of measurement leads to a distribution of rupture forces. In addition, measurements of rupture forces over a wide range of pulling velocities, from very slow to extremely fast, are used to explore the energy landscape of the bound complex.

In DFS experiments an adhesion bond is driven away from its equilibrium by a spring pulled at a given velocity. Rupture of adhesion bonds occurs via thermally assisted escape from the bound state across an activation barrier. The latter diminishes as the applied force increases, so the rupture force is determined by an interplay between the rate of escape in the absence of the external force, and the pulling velocity (loading rate).

Thus, the measured forces are not an intrinsic property of the bound complex, but rather depend on the mechanical setup and loading rate applied to the system.

In the present report we discuss a new approach to describe unbinding processes measured by DFS, which goes beyond previous models and methods of analysis [2-10]. As we show, our approach: (1) proposes a possible mechanism of rupture, (2) emphasizes the importance of investigating the *distribution* function of rupture forces rather than focusing on *typical* rupture forces only, (3) includes the dynamics of conformational degrees of freedom, and (4) gives a deeper insight into the effects of rebinding.

**Single-path picture.** Let us start from a one-dimensional description of the unbinding process along a single reaction coordinate, $x$. The dynamic response of the bound complex is governed by the Langevin equation

$$M \ddot{x}(t) = -\boldsymbol{g}_x \dot{x}(t) - \frac{\partial U(x)}{\partial x} - K(x - Vt) + \boldsymbol{x}_x(t). \qquad (1)$$

Here the molecule of mass $M$ is pulled by a linker of a spring constant $K$ moving at a velocity $V$. $U(x)$ is the adhesion potential, $\boldsymbol{g}_x$ is a dissipation constant and the effect of thermal fluctuations is given by a random force $\boldsymbol{x}_x(t)$, which is $\boldsymbol{d}$-correlated $<\boldsymbol{x}_x(t)\boldsymbol{x}_x(0)> = 2k_B T \boldsymbol{g}_x \boldsymbol{d}(t)$. In Eq.(1) thermal fluctuations are the origin of the distribution of rupture forces. In a more general case other sources of randomness are possible.

The bound state is defined by the minimum of the *total* potential $\boldsymbol{F}(X,t) = U(x) + \frac{K}{2}(x-Vt)^2$. In the absence of thermal fluctuations unbinding occurs when the potential barrier vanishes, i.e. at the instability point where $d^2\boldsymbol{F}(x,t)/dx^2=0$, $d\boldsymbol{F}(x,t)/dx=0$. At this point the measured spring force, $F=K(x-Vt)$, reaches its maximum value $F=F_c$. In the presence of fluctuations the escape from the potential well occurs earlier, and the probability $W(t)$ that a molecule persists in its bound state is defined by Kramer's transition rate [11] and can be approximately calculated through the following kinetic equation

$$\frac{dW(t)}{dt} = -\frac{\Omega_1(t)\Omega_2(t)}{2\pi g_x M}\exp[-\Delta E(t)/k_B T]W(t). \qquad (2)$$

Here $\Delta E(t)$ is the instantaneous barrier height and $\Omega_{1,2}(t)$ are the effective oscillation frequencies at the minimum corresponding to the bound state and maximum of the combined potential $F(x,t)$. Equation (2) does not take into account rebinding processes. The experimentally measured distribution of rupture forces, $P(F_{max})$, can be expressed in terms of $W$ as

$$P(F_{max}) = -\frac{d}{dF_{max}}W(F_{max}), \text{ and } \langle F_{max} \rangle = -\int_0^\infty F'_{max}\left(\frac{d}{dF'_{max}}W(F'_{max})\right)dF'_{max}. \qquad (3)$$

As we have already noted above the rupture force, $F_{max}$, is defined as the maximal spring force, $K(x-Vt)$, measured during rebinding process.

Because of the exponential dependence of the unbinding rate on $\Delta E(t)$, we focus on values of $F$ close to the critical force $F_c$ at which the barrier disappears completely. Then the instantaneous barrier height and the oscillation frequencies can be written in terms of the reduced bias [12], $\varepsilon = 1 - F_{max}/F_c$, as

$$\Delta E(t) = U_c \varepsilon^{3/2}, \qquad \Omega_{1,2}(t) = \Omega_c \varepsilon^{1/4}, \qquad (4)$$

where $U_c$ and $W_c$ are the parameters of the bare, unbiased, potential $U(x)$ which is an *information we are after*. See ref. [13] for the definitions of $U_c$, $W_c$ and $F_c$ corresponding to Morse potential.

Solution of the kinetic equation (2) with $\Delta E(t)$ and $\Omega_{1,2}(t)$ given by Eq.(4) leads to the final expressions for $P(F_{max})$, its mean value $<F_{max}>$, and the variance $\sigma^2_{F_{max}} = \langle F^2_{max} \rangle - \langle F_{max} \rangle^2$:

$$\langle F_{max} \rangle \approx F_c \left\{ 1 - \left( \frac{k_B T}{U_c} \right)^{2/3} \left[ \ln\left( \frac{U_c 3\pi g_x K}{k_B T \Omega_c^2 M F_c} V \right) \right]^{2/3} \right\}; \quad (5)$$

$$P(F_{max}) = P_0 \varepsilon^{1/2} \exp\left\{ -\frac{U_c}{k_B T} \varepsilon^{3/2} - \frac{k_B T \Omega_c^2 M F_c}{U_c 3\pi g_x K V} e^{-\frac{U_c}{k_B T} \varepsilon^{3/2}} \right\}; \quad (6)$$

$$\sigma^2_{F_{max}} = \frac{2\pi^2}{27} F_c^2 \left( \frac{k_B T}{U_c} \right)^{4/3} \left[ \ln\left( \frac{F_c \Omega_c^2 M k_B T}{KV 3\pi g_x U_c} \right) \right]^{-2/3}; \quad (7)$$

where $P_0$ is a normalization constant.

Expression (5) differs essentially from the earlier proposed and often used, logarithmic law, $<F_{max}> = const + (k_B T / \Delta x) \ln[VK\Delta x/(k_0 k_B T)]$ [3,5], where $k_0$ is the spontaneous rate of bond dissociation, and $\Delta x$ is a distance from the minimum to the activation barrier of the reaction potential $U(x)$. The logarithmic law has been derived within a Kramers picture for the escape from a well (bound state) assuming that the pulling force produces a small constant bias which reduces the height of a potential barrier. This is, however, an unlikely regime. As the linker is driven out of the adhesion complex and the bias is ramped up, a bond rupture occurs preferentially when a potential barrier almost vanishes. Similar mechanism has been recently suggested for interpretation of the effect of thermal fluctuations on atomic friction [14, 15].

**Scaling of rupture forces.** Equation (5) predicts a universal scaling, independent of temperature, of $\left( F_c - \langle F_{max} \rangle \right)^{3/2} / T$ with $\ln(V/T)$. Fig.1A shows the agreement between numerical calculations using Langevin Eq.(1) and the analytical form in Eq.(5). Over a wide range of pulling velocities the numerical data obtained for three different temperatures collapse on a single straight line when plotted as $\left(F_c - \langle F_{max} \rangle\right)^{3/2}/T$ vs $\ln(V/T)$. In contrast, when examining the expression $\langle F_{max} \rangle \propto \ln(V/T)$ the scaling breaks down (see inset to Fig.1A). The proposed scaling can be tested experimentally for unbinding and has been shown to work in friction experiments [15]. Fig.1B displays numerical results in agreement with the form in Eq.(6) for the distribution function of rupture forces calculated for a given velocity. We note the non-Gaussian nature of the distribution and its pronounced asymmetry. Such an asymmetry has been reported already for both small molecules [16] and macromolecules [17]. The width of the distribution is given by $s^2_{F_{max}}$ and shows a decrease with a decrease in pulling velocity $V$.

Our results suggest that fitting experimental data to Eqs.(5) and (6) one can determine three microscopic parameters of the adhesive potential: $F_c$, $U_c$ and $(U_c 3 p g_x K)/(\Omega_c^2 F_c M)$. Thus, DFS experiments can provide a new complementary information on adhesive potentials when compared to equilibrium measurements which provide the spontaneous rate of bond dissociation.

According to Fig.1A the numerical data at very low and very high pulling velocities deviate from the predicted straight line, $\left(F_c - \langle F_{max} \rangle\right)^{3/2}/T$ vs $\ln(V/T)$. The deviation at high driving velocities results from the dominating effect of viscous dissipation given by the term $g_x \dot{x}$ in Eq.(1), which is not included in the kinetic model, Eq.(2). The deviation of the unbinding force from the analytical form (5) at low pulling velocities is a direct result of rebinding events discussed by Evans [3], Seifert [9] and

Prechtel *et al* [18], and which is clearly observed in the time series of the spring force shown in Fig.1C. The rebinding appears as an intermittent series resembling stick-slip motion in friction measurements. The deviation of the high temperature curve from scaling in Fig.1A marks the setting in of rebinding at this temperature.

**Beyond the single-path picture.** The above considerations are based on the assumption that unbinding occurs along a single path within the energy landscape. However, due to the multidimensionality of the energy landscape of complex molecules, a distribution of configurations might be involved which depends on the applied force. Due to finite relaxation times, such configurations cannot always follow spontaneously the applied force. This could lead to a richer dependence of the rupture force on the pulling velocity. The unbinding process is now determined by the motion in the space of at least two "reaction" coordinates: the coordinate $x$ of the linker, which provides the applied force, and a "collective" coordinate $f$, which characterizes the configurational state of the molecule. Thus, the adhesion potential $U(x,f)$ is a function of these two coordinates. The system chooses an optimal pathway in the space of these two variables. As we show below, the configurational state of the molecule at the moment of rupture can depend on the driving velocity that leads to a new behavior of the rupture forces. To describe a dynamics of configurational state, we introduce, in addition to Eq.(1), an equation for the "collective" coordinate,

$$g_f \dot{f}(t) = -\frac{\partial U(x,f)}{\partial f} + \xi_f(t), \qquad (8)$$

where the dissipation constant $g_f$ defines a timescale of the configurational relaxation and $<\xi_f(t)\xi_f(0)> = 2k_B T g_f \delta(t)$.

Under equilibrium conditions, and within a two-state picture, the molecule passes from a "folded" to an "unfolded" configuration with an increase in the external force. A region of bistability, where both configurational states are populated, has been actually observed in experiments where a constant force has been applied [19]. These observations indicate that for a given position $x$ of the linker, the potential $U(x,f)$ as a function of the configuration coordinate $f$ has two minima. For a small spring extension the population of configurational states is localized mostly in the vicinity of the "folded" minimum of $U$, while for larger applied forces it is expected to be localized around the "unfolded" minimum (see Fig.2). This behavior reflects a tilting of the folded-unfolded equilibrium toward the unfolded state with an increase of the force.

Assuming that the conformational coordinate $f$ represents a growing molecular size (for instance, end-to-end distance), we take $U(x, f)$ as

$$U(x,f) = U_f(f) + U_{un}(f - x), \qquad (9)$$

where $U_f$ and $U_{un}$ are each a single-well potential with minima, $f = f_f$ and $f = f_{un} + x$ corresponding to the folded and unfolded configurations, respectively (see Fig.2). The $f$-position of the second minimum grows with $x$, reflecting the fact that in the unfolded configuration the size of the molecule follows the position of the linker, $x$. The forces experienced by the linker during an unbinding process are different in the two configurational states [20]. It is natural to assume (see also experiments [19]) that for a given extension $x$ of the linker, the force in the folded configuration is larger than in the unfolded one.

For a very slow increase of the external force, where the system remains close to equilibrium, the populations of the folded and unfolded states are determined by the depths of the corresponding potential wells. Rupture usually occurs in the unfolded state. The situation is different in experiments where the applied force increases with a finite rate. Our calculations demonstrate that due to the finite relaxation time of the molecular degrees of freedom, the populations depend on the driving velocity. With an increase in velocity ($V>a/?_f$, $a$ being a constant) the configurational population is not able to follow the applied force, and rupture occurs in the folded configuration. This behavior is clearly seen in Fig.3 showing an evolution of the rupture force with pulling velocity, and the distribution functions of rupture forces for three chosen velocities. The bottom branch of the force corresponds to the rupture occurring in the unfolded configuration while the upper branch to the rupture in the folded configuration. Fig.3 demonstrates that there is a region of velocities where rupture can occur in both folded and unfolded configurations.

Analyzing the two branches of the force within the single path approach, one can extract information on the two configurational state potentials. In addition, one can obtain information on the time $1/g_f$ for the transition between the folded and unfolded configurations. This can be estimated from the value of velocity at the bottom boundary of the bistability region.

The authors thank M. Elbaum, R. Nevo and Z. Reich for useful discussions.

Financial support for this work by grants from the Israel Science Foundation, BSF, TMR-SISITOMAS is gratefully acknowledged.

20. In the folded state, $f = f_f$, $F = -\left(\partial U(x,\mathbf{j})/\partial x\right)_{\mathbf{j}=\mathbf{j}_f} = -\partial U_{un}(\mathbf{j}_f - x)/\partial x$ while in the unfolded state $F = -\left(\partial U(x,\mathbf{j})/\partial x\right)_{\mathbf{j}=x+\mathbf{j}_{un}} = -\partial U_f(x+\mathbf{j}_{un})/\partial x$.

## Figure captions.

Fig.1. Analysis of rupture process for a single reaction coordinate. **1A** Results of numerical calculations supporting the scaling behavior of the ensemble averaged rupture force according Eq.(5). The inset shows a significantly worse scaling for the description $<F_{max}> \propto const - \ln(V/T)$. The units of velocity $V$ are nm/sec, temperature is in degrees kelvin. Analyzing the numerical data presented in Fig.1 gives the following values: $F_c=0.77$ nN, $U_c=0.12$ nN nm, and $(U_c 3\mathbf{pg}_x K)/(\Omega_c^2 F_c M) = 6.8 \times 10^{-7}$ nN sec. This compares well to the corresponding values of $F_c=0.75$ nN, $U_c=0.12$ nN nm, and $(U_c 3\mathbf{pg}_x K)/(\Omega_c^2 F_c M) = 6.8 \times 10^{-7}$ nN sec used in our numerical calculations. **1B** Normalized distribution of the unbinding force at temperature $T=293$K for two values of the velocity. The result from the numerical simulation (solid line) is in a good agreement with the theoretical distribution, Eq.(6), for velocity $V=117$ nm/c. For velocity $V=5.9$ nm/sec, where the rebinding plays an essential role, the distribution

function deviates from the one given by Eq.(6). **1C** Time series of the spring force showing the rebinding events for $T=293$K, $V=5.9$ nm/sec. Parameter values: $K=0.93$ N/m, $\gamma=7.7\times10^{-6}$ sec$^{-1}$ kg, $M = 8.7\times10^{-12}$ kg, $U_0 = 0.12$ nN $\times$ nm, $R_c=0.24$ nm, $b=1.5$.

Fig.2. Schematic presentation of the typical stages of evolution of the adhesion potential $U(x,f)$ and the corresponding distribution function in configurational space, $f$, with variation of the linker position under quasi-equilibrium conditions: **A** zero applied force, **B** low applied force, **C** high applied force. Bold and thin lines show the configurational populations and potential correspondingly. Left (right) potential minimum corresponds to the folded (unfolded) state.

Fig.3. Probability density of the rupture force, $P(F_{max})$. The top panel shows the velocity dependence of $P(F_{max})$, red and blue color correspond to the maxima and minima of $P(F_{max})$. The bottom panel shows force histograms corresponding to three typical velocities indicated by arrows in the top panel. The potential (9) is approximated by the expression $U_{f,un} = -\frac{1}{2}C_{f,un}e^{-(j/c_{f,un})^2} + D_{f,un}e^{-j/d_{f,un}}$. Parameter values: $c_f/c_{un}=0.68$, $d_f/c_{un}=0.06$, $c_{un}/c_{un}=0.25$, $C_f/C_{un}=0.61$, $D_f/C_{un}= D_{un}/C_{un}=11.56$, $Kc_{un}^2/C_{un} = 0.0075$, $V\gamma c_{un}/C_{un} = V\cdot 0.0093$. Forces and velocities are in units of $C_{un}/c_{un}$ and $C_{un}/(\gamma c_{un})$ respectively.

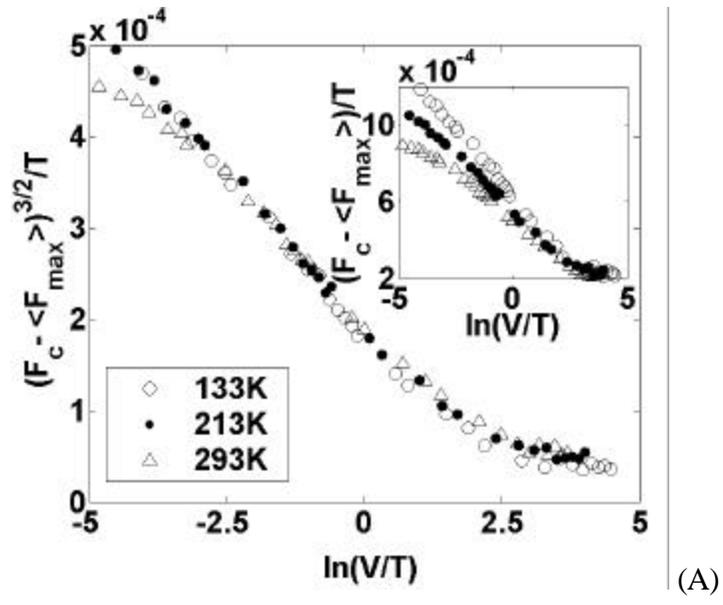

(A)

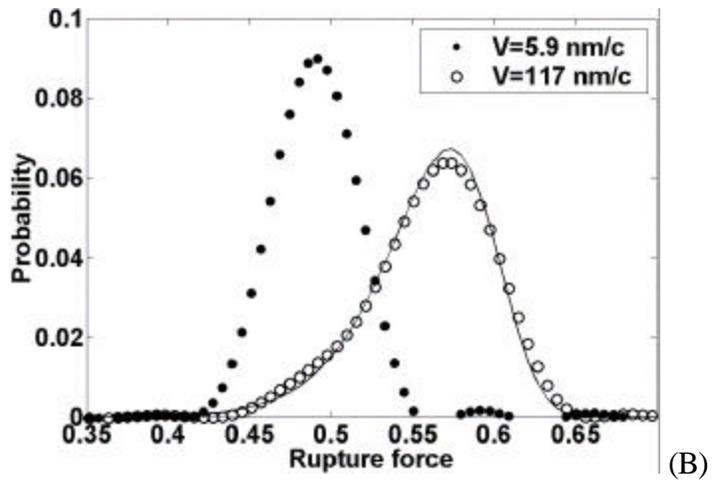

(B)

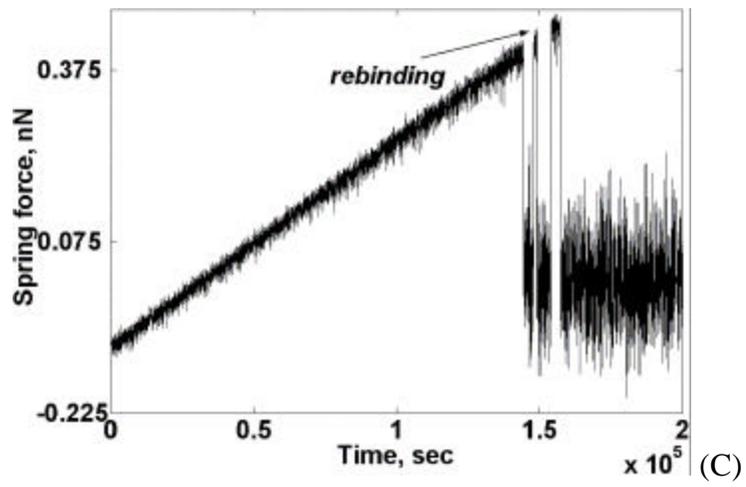

(C)

Fig. 1

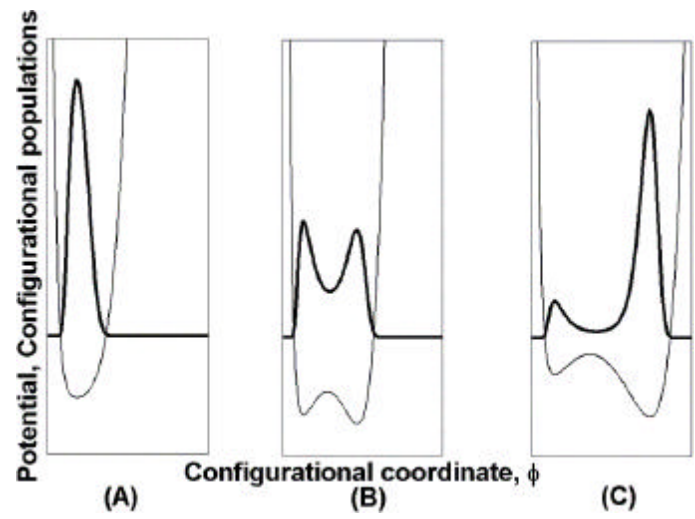

Fig.2

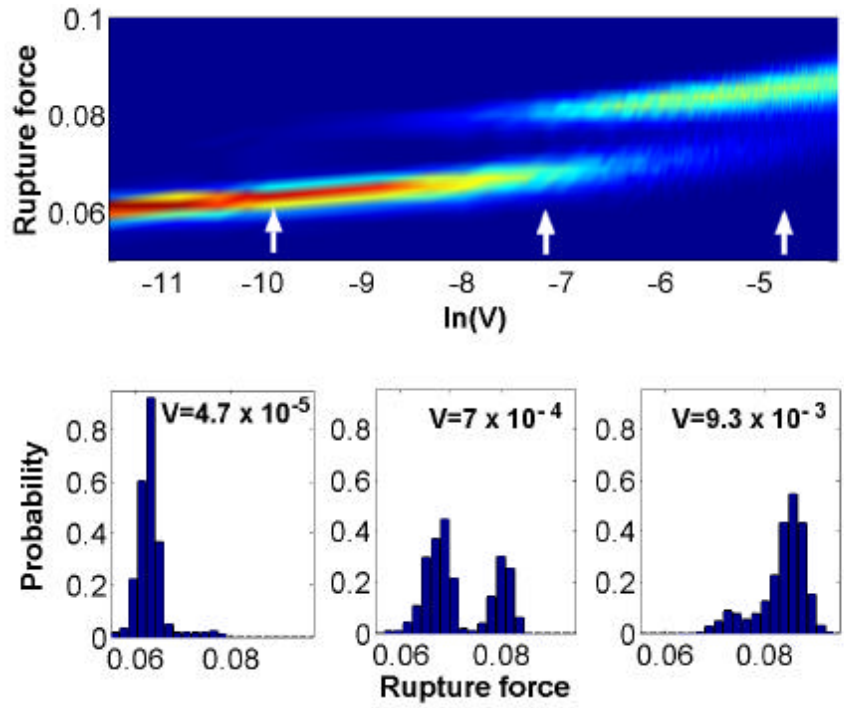

Fig.3